\documentclass[
  ,draft            
  ]
  {aipproc}

\layoutstyle{6x9}

\begin{document}

\title{$V_{us}$ From Hadronic $\tau$ Decays}


\author{K. Maltman}{
  address={Dept. Math and Stats, York Univ., 4700 Keele St., Toronto,
ON CANADA M3J 1P3 and CSSM, Univ. of Adelaide, Adelaide 5005 SA, Australia}}

\author{C.E. Wolfe}{
  address={Dept. Physics and Astronomy, York Univ., 4700 Keele St., Toronto,
ON CANADA M3J 1P3}}

\begin{abstract}
We study extractions of $\vert V_{us}\vert$ based on finite
energy sum rule (FESR) analyses of hadronic $\tau$ decay
data. We show (i) that the ``$(0,0)$ spectral weight'' implementation
(proposed previously in the literature as a favorable version
of this analysis) suffers from significant convergence problems, but
(ii) that alternate implementations exist which bring these problems 
under control. Results based on present spectral data are shown 
to be in agreement with those of other approaches, though with somewhat
larger experimental errors. Sub-$1\%$ determinations of $\vert V_{us}\vert$
are also shown to be possible once $\tau$ data from the B 
factories becomes available.
\end{abstract}

\maketitle


There has been considerable recent interest in the
status of the first row CKM unitarity relation,
$\vert V_{ud}\vert^2+\vert V_{us}\vert^2+\vert V_{ub}\vert^2=1$.
Recent determinations of $\vert V_{us}\vert$ from
(i) $\Gamma_{K_{\mu 2}}/ \Gamma_{\pi_{\mu 2}}$, analyzed
using lattice results for $f_K/f_\pi$ as input~\cite{marcianoetc,ckmwg}
and (ii) $K_{\ell 3}$ data, using updated results for the 
kaon lifetimes, partial widths and $K_{\ell 3}$ form factor slope parameters, 
combined with lattice and ChPT work on the deviation of $f^{K\pi}_+(0)$ 
from $1$~\cite{kell3stuff}, yield results compatible, within errors, 
with 3-family unitarity. A major component of the
error in both approaches is theoretical, and associated
with uncertainties in the extrapolation of lattice data to physical
values of the light quark masses. Here we discuss an
alternate approach to extracting $\vert V_{us}\vert$, based on
analyses of flavor-breaking sum rules involving 
hadronic $\tau$ decay data~\cite{gamizetal,kmcw06}. 
An important feature of this method is that the associated 
theoretical errors are completely independent 
of the chiral extrapolation uncertainties which dominate the 
$K_{\ell 3}$ and $K_{\mu 2}$ vs. $\pi_{\mu 2}$ approaches.

The $\tau$-based method works as follows.
With $\Pi^{(J)}_{V/A;ij}$ the spin $J$ parts of the flavor
$ij=ud,us$ vector/axial vector correlators, $\rho^{(J)}_{V/A;ij}$
the corresponding spectral functions, and 
$R_{V/A;ij}\, =\, \Gamma [\tau\rightarrow \nu_\tau\, {\rm hadrons}_{V/A;ij}
\, (\gamma )]/\Gamma [\tau\rightarrow \nu_\tau \bar{\nu}_e e^-\, (\gamma )]$,
$\tau$ decay kinematics yield
\begin{equation}
R_{V/A;ij}= 12\pi^2\vert V_{ij}\vert^2 S_{EW}\, \int^1_0 dy_\tau\,
\left[ w_T(y_\tau )\rho^{(0+1)}(s)\, +\, w_L(y_\tau )\rho^{(0)}(s)\right]\ ,
\label{basictauspeccombo}\end{equation}
where $y_\tau =s/m_\tau^2$, $S_{EW}$ is a short-distance electroweak
correction, $w_T(y)\, =\, (1-y)^2(1+2y)$, $w_L(y)\, =\, -2y(1-y)^2$,
and the superscript $(0+1)$ denotes the sum of $J=0$ and $J=1$
contributions. As written, the RHS of Eq.~(\ref{basictauspeccombo}) 
can be recast using the general FESR relation,
$\int_0^{s_0}ds\, w(s)\rho (s)\, =\, {\frac{-1}{2\pi i}}\,
\oint_{\vert s\vert =s_0}ds\, w(s)\Pi (s)$,
valid for any $\Pi$ without kinematic singularities and any
analytic $w(s)$. Rescaling the experimental decay distribution in 
Eq.~(\ref{basictauspeccombo}) by $(1-y_\tau )^k y_\tau^m$ ($k,m\geq 0$)
before integration leads to the generalized spectral integrals,
$R_{V/A;ij}^{(k,m)}$, and corresponding ``(k,m) spectral weight sum rules''.
Analogous spectral integrals and FESR's can also be
written down for $s_0<m_\tau^2$, for the separate correlator combinations
$\Pi^{(0+1)}_{V/A;ij}(s)$, $s\Pi^{(0)}_{V/A;ij}(s)$, and for general 
$w(s)$. The corresponding spectral
integrals are denoted generically $R^{w}_{ij}(s_0)$ below. At
sufficiently large $s_0$ the OPE representation
can be employed for $\Pi (s)$ on the RHS of the FESR relation.

With this notation, choosing a particular $s_0$, $w(s)$
and correlator combination, the spectral integral difference
$\delta R^w(s_0)\, =\, \left[ R^w_{ud}(s_0)/\vert V_{ud}\vert^2\right]
\, -\, \left[ R^w_{us}(s_0)/\vert V_{us}\vert^2\right]$
vanishes in the SU(3) flavor limit. Its OPE representation,
$\delta R^w_{OPE}(s_0)$, thus begins at dimension $D=2$. Using this 
representation, and solving for $\vert V_{us}\vert$, one has~\cite{gamizetal}
\begin{equation}
\vert V_{us}\vert\, =\, \sqrt{R^w_{us}(s_0)/
\left[\left( R^w_{ud}(s_0)/\vert V_{ud}\vert^2\right)\, 
-\, \delta R^w_{OPE}(s_0)\right]}\ .
\end{equation}
An important practical observation is that, 
for a range of weights studied in the literature,
$\delta R^w_{OPE}(s_0)$ is at the few to several $\%$ level of 
$R^w_{ud}(s_0)/\vert V_{ud}\vert^2$. High-precision determinations 
of $\vert V_{us}\vert$ are thus possible from modest-precision determinations 
of $\delta R^w_{OPE}(s_0)$, provided errors on the $ud$ and $us$
spectral integrals are kept under control~\cite{gamizetal}.
The following technical points are relevant to constructing reliable 
versions of this analysis:
\begin{itemize}
\item Because (i) the integrated $D=2$ OPE for the purely
$J=0$ contributions displays extremely poor convergence for all $s_0$
accessible in hadronic $\tau$ decay~\cite{convergence}, and 
(ii) all truncation schemes considered
in the literature produce strong violations of spectral 
positivity~\cite{positivity}, it is necessary to work with
FESR's based on the difference, $\Delta\Pi$, of the
$ud$ and $us$ $J=0+1$ correlators.
Fortunately, for a combination of chiral and kinematic reasons,
the $J=0$ contributions to the experimental spectral distribution
are strongly dominated by the well-known $K$ and $\pi$ pole
terms, and hence can be straightforwardly subtracted. 
The desired $J=0+1$ spectral difference is then easily determined. 
(Small residual corrections are constrained by analyses of
the strange scalar and pseudoscalar sectors~\cite{longcontinuum}.)
\item The series in $\alpha_s$ for $\left[\Delta\Pi\right]^{OPE}_{D=2}$ 
is slowly converging at the spacelike point on $\vert s\vert=s_0$, 
for all $s_0<m_\tau^2$~\cite{bck05,footnote1}. This necessitates
using weights which emphasize other regions of the
contour, where $\vert \alpha_s (Q^2)\vert$ 
is smaller and the convergence of the correlator series
improved. The $(k,0)$ spectral weights are disfavored from this 
point of view since the modulus of $(1-y)^{k+2}$ peaks more
and more strongly in the spacelike direction ($y\, =\, -1$)
as $k$ is increased. The convergence
of the integrated $D=2$ OPE series is already very poor for the
$(0,0)$ case~\cite{bck05,kmjk00}. Fortunately, alternate weights,
$w_{10}$, $\hat{w}_{10}$ and $w_{20}$~\cite{kmjk00},
displaying significantly improved convergence behavior~\cite{kmcw06,kmjk00},
exist in the literature.
\item Lack of empirical information on higher $D$ VEV's
necessitates neglect of $D>6$ or $8$ OPE contributions.
This neglect can be tested for self-consistency by 
employing weights $w(y)$, with $y=s/s_0$,
and checking output $\vert V_{us}\vert$ values for any unphysical 
$s_0$-dependence, since integrated $D=2k$ contributions then 
scale as $1/s_0^k$ relative to the leading integrated $D=2$ term.
Such unphysical $s_0$-dependence can also result from
truncation of a slowly converging integrated $D=2$ series. Testing the output
for $s_0$-stability is thus {\it very} important,
especially when, as for the $(0,0)$ spectral
weight, convergence is slow and conventional truncation error 
estimates, as a result, are potentially suspect.
{\it To make such $s_0$-stability tests possible,
it is essential that experimentalists provide
full covariance matrices, and not just quote correlated
errors for a limited number of weights and for $s_0=m_\tau^2$
only.} 
\end{itemize}

With currently available CLEO, ALEPH and OPAL data~\cite{spectraldata},
$ud$ and $us$ spectral integral errors ($\sim 1/\%$ and $\sim 3-4\%$,
respectively) lead to $\sim 1/4\%$ and $\sim 1.5-2\%$ errors on 
$\vert V_{us}\vert$. We have employed
the ALEPH data and covariance matrices, with 
the distribution for each exclusive mode rescaled by the ratio
of the current world average to original ALEPH branching fraction for that 
mode~\cite{chen01}. The largest rescaling occurs for the $K^\pm \pi^+\pi^-$
mode, where the OPAL and CLEO results are in good agreement,
but in disagreement with ALEPH. We quote results below for both
the 3-fold ALEPH-CLEO-OPAL ('ACO') and 2-fold CLEO-OPAL ('CO') averages.
The dominant $us$ spectral integral errors will be dramatically 
reduced once analyses
of the much larger data sets available from the B factory experiments
become available.

OPE errors are dominated by uncertainties in the leading integrated 
$D=2$ series. The results reported below correspond to the PDG04 input, 
$m_s(2\ {\rm GeV})=105\pm 25\ {\rm MeV}$, whose square
sets the overall scale for the $D=2$ series. Other OPE input is
detailed in Ref.~\cite{kmcw06}. The $D=2$ error has two main sources: 
the truncation of the $D=2$ series, 
and the overall $\left[ m_s(2 \ {\rm GeV})\right]^2$ scale uncertainty.
The latter can be easily reduced through near-term improvements in 
our knowledge of $m_s$, while the former are not subject to near-term
improvement. A useful indicator for the quality of the convergence
of the integrated $D=2$ series is $r_k(s_0)$, the difference between
the $O(\alpha_s^k)$-truncated correlator and Adler function versions of the 
$D=2$ integral, scaled by the corresponding correlator integral.
The two expressions must be identical to all orders but, at fixed order 
$k$, differ by terms of $O(\alpha_s^{k+1})$. For a well-converged series, 
$r_k(s_0)$ should be small, and decreasing in magnitude with increasing $k$.
For $w_{10}$, $\hat{w}_{10}$ and $w_{20}$, we find
$r_{2,3,4}(m_\tau^2)\, =\, -0.06,\, -0.03$, $-0.01$; 
$r_{2,3,4}(m_\tau^2)\, =\, -0.07,\, -0.05$, $-0.03$; and
$r_{2,3,4}(m_\tau^2)\, =\, -0.08,\, -0.05$, $-0.03$, respectively, 
confirming the (by design) improved convergence of the corresponding 
integrated $D=2$ series. For the $(0,0)$ spectral weight, in contrast, 
$r_{2,3,4}(m_\tau^2)\, =\, 0.06,\, 0.20$, $0.67$,
confirming the poor convergence of the integrated $D=2$ series.
The latter result raises warning flags about the reliability 
of existing $(0,0)$ spectral weight truncation 
error estimates (the difference between
the correlator and Adler function evaluations alone is a factor
of $\sim 2.5$ times larger than the original estimate
of Refs.~\cite{gamizetal}).
$D=2$ errors turn out to be dominated by the (sizeable) truncation
uncertainty (and hence not improvable) for the $(0,0)$ spectral weight, 
but by the overall scale uncertainty (and hence easily improvable) for
$w_{10}$, $\hat{w}_{10}$ and $w_{20}$~\cite{kmcw06}. A more detailed breakdown
and discussion of OPE errors can be found in Ref.~\cite{kmcw06}.

The results of the $s_0$-stability tests for the various weights
are displayed in Table~\ref{table1}. One sees poor
stability, and hence significant additional theoretical uncertainty
(much greater than the total $\pm 0.0009$ theoretical uncertaintiy
estimated in Refs.~\cite{gamizetal}),
for the $(0,0)$ spectral weight, but good stability 
for $w_{10}$, $\hat{w}_{10}$ and $w_{20}$.

\begin{table}[ht]
\begin{tabular}{|l|c|ccc||c|ccc|}
\hline
$s_0$&$w^{(0,0)}_{ACO}$&$\hat{w}^{ACO}_{10}$&$w^{ACO}_{20}$&$w^{ACO}_{10}$&
$w^{(0,0)}_{CO}$&$\hat{w}^{CO}_{10}$&$w^{CO}_{20}$&$w^{CO}_{10}$\\
\hline
2.35&0.2149&0.2220&0.2243&0.2201&0.2172&0.2236&0.2255&0.2218\\
2.55&0.2167&0.2218&0.2235&0.2203&0.2192&0.2236&0.2250&0.2223\\
2.75&0.2181&0.2218&0.2230&0.2207&0.2207&0.2239&0.2246&0.2229\\
2.95&0.2193&0.2220&0.2227&0.2211&0.2219&0.2243&0.2245&0.2235\\
3.15&0.2202&0.2223&0.2226&0.2216&0.2228&0.2246&0.2246&0.2241\\
\hline
\end{tabular}
\caption{$\vert V_{us}\vert$ as a function
of $s_0$ for various FESR weight choices and the ACO and CO treatments of the 
$us$ data. $s_0$ is given in units of GeV$^2$}\label{table1}
\end{table}

Our final results are based on the $\hat{w}_{10}$ and $w_{20}$
weights, and $s_0=m_\tau^2$, for reasons discussed 
in Ref.~\cite{kmcw06}. We find, for the ACO (CO) treatments of the
$us$ data,
\begin{equation}
\vert V_{us}\vert\ =\ 0.2223\ (0.2246)\ \pm 0.0032_{exp}\pm 0.0038_{th}\ ,
\end{equation}
both results in agreement, within errors, with those of 
other methods. We stress that 

(i) the errors here are entirely
independent of those of the other methods; 

(ii) the experimental
errors will be subject to significant near-term improvement once
data from the B factory experiments becomes available; 

(iii) the theoretical errors for the $(0,0)$ spectral weight
approach are dominated by $D=2$ convergence problems, and
not subject to future improvement, making
this version of the analysis unfavorable; and

(iv) for the alternate weights, 
the theoretical errors are dominated by the $D=2$ 
scale uncertainty and reducible to the $\pm 0.0013$
level (or below) once the uncertainty in $m_s(2\ {\rm GeV})$ is
reduced to $\pm 5\ {\rm MeV}$~\cite{kmcw06}. 
Sub-$1\%$ determinations
of $\vert V_{us}\vert$ from hadronic $\tau$
decay analyses are thus a realistic goal, with a further reduction
in errors possible through averaging with the output from other methods.






\bibliographystyle{aipproc}   




\end{document}